\documentclass[11pt]{llncs}

\usepackage{amsfonts}

\newcommand{\realrange}[2]{\left[#1, #2\right]}

\newcommand{\unitrange}[2]{\realrange{0}{1}}

\newcommand{\Oh}[1]{\mathcal{O}\!\left( #1\right)}

\newcommand{\llabel}[1]{\label{\labelprefix:#1}}
\newcommand{\labelprefix}{} 

\newcommand{\discussionsize}{\small}

\newcommand{\frage}[1]{}

\newenvironment{code}{\noindent
\begin{tabbing}%
\hspace{2em}\=\hspace{2em}\=\hspace{2em}\=\hspace{2em}\=\hspace{2em}\=%
\hspace{2em}\=\hspace{2em}\=\hspace{2em}\=\hspace{2em}\=\hspace{2em}\=%
\kill}{\end{tabbing}}

\newcommand{\labelcommand}{}
\newcommand{\captiontext}{}
\newsavebox{\codeparam}
\newcounter{lineNumber}
\newenvironment{disscodepos}[3]{%
\renewcommand{\labelcommand}{#2}%
\renewcommand{\captiontext}{#3}%
\sbox{\codeparam}{\parbox{\textwidth}{#3}}%
\begin{figure}[#1]\begin{center}\begin{code}\setcounter{lineNumber}{1}}{%
\end{code}\end{center}\caption{\llabel{\labelcommand}\captiontext}\end{figure}}

{\end{disscodepos}}

\newdimen\endofsize\endofsize=0.5em
\def\endofbeweis{~\quad\hglue\hsize minus\hsize
                 \hbox{\vrule height \endofsize width
\endofsize}\par}

\usepackage{fullpage}     
\usepackage{times}     
\usepackage{multirow}
\usepackage{longtable}
\usepackage{numprint}
\usepackage{wrapfig}
\usepackage{amsmath}
\usepackage{hyperref}
\usepackage[usenames,dvipsnames]{xcolor}
\usepackage[normalem]{ulem}
\usepackage[]{graphicx}\usepackage[]{color}
\makeatletter
\def\maxwidth{ %
  \ifdim\Gin@nat@width>\linewidth
    \linewidth
  \else
    \Gin@nat@width
  \fi
}
\makeatother

\definecolor{fgcolor}{rgb}{0.345, 0.345, 0.345}

\usepackage{framed}
\makeatletter
 {\par\unskip\endMakeFramed%
 \at@end@of@kframe}
\makeatother

\definecolor{shadecolor}{rgb}{.97, .97, .97}
\definecolor{messagecolor}{rgb}{0, 0, 0}
\definecolor{warningcolor}{rgb}{1, 0, 1}
\definecolor{errorcolor}{rgb}{1, 0, 0}
\newenvironment{knitrout}{}{} 

\usepackage{alltt}
\newcommand{\SweaveOpts}[1]{}  
\newcommand{\SweaveInput}[1]{} 
\newcommand{\Sexpr}[1]{}       

\usepackage{booktabs}

\newcommand{\ie}{i.e.~}
\newcommand{\etal}{et~al.\ }
\newcommand{\eg}{e.g.\ }

\usepackage{capt-of}%
\usepackage{caption}

\usepackage{etoolbox}
\usepackage[left,pagewise] {lineno}
\definecolor {infocolor} {rgb} {0.6,0.6,0.6}

\patchcmd{\thebibliography}{\list}{\fontsize{0.98em}{0.9\baselineskip}\selectfont\list}{}{} 

\newcommand{\mytitle}{Memetic Multilevel Hypergraph Partitioning}
\begin{document}
\title{\mytitle}
\author{Robin Andre, Sebastian Schlag and Christian Schulz \\
        \vspace*{.25cm}
\textit{Karlsruhe Institute of Technology},
  \textit{Karlsruhe, Germany} \\
  \small	\textit{\url{robin.andre@ira.uka.de}, \{\url{sebastian.schlag, christian.schulz}}\}\url{@kit.edu}  \\
        \vspace*{.25cm}
  \normalsize        \textit{University of Vienna}, \textit{Vienna, Austria} \\
  \small	\textit{\url{christian.schulz}\url{@univie.ac.at}}
} 

\date{}
\institute{}

\maketitle
\begin{abstract}
Hypergraph partitioning has a wide range of important applications such as VLSI design or scientific computing.  With focus on solution quality, we develop the first multilevel memetic algorithm to tackle the problem. Key components of our contribution are new effective multilevel recombination and mutation operations that provide a large amount of diversity. We perform a wide range of experiments on a benchmark set containing instances from application areas such VLSI, SAT solving, social networks, and scientific computing. Compared to the state-of-the-art hypergraph partitioning tools hMetis, PaToH, and KaHyPar, our new algorithm computes the best result on almost all instances.
\end{abstract}
\thispagestyle{empty}
\vfill
\pagebreak
\section{Introduction}
Given an undirected hypergraph $H=(V,E)$, the \emph{$k$-way hypergraph partitioning problem} 
is to find disjoint subsets of its vertex set, $V_1, \dots, V_k$, called \emph{blocks}, such that the
blocks have roughly equal size and an objective function involving the cut hyperedges is minimized, \eg, the sum of the weights of those hyperedges that connect multiple blocks. 
The hypergraph partitioning problem has many important applications in practice such as  
scientific computing or VLSI design~\cite{Papa2007}.
In particular VLSI design is a field where small improvements can lead to significant savings~\cite{Wichlund98}.
Hence, our focus in this work is on \emph{solution quality}.

It is well known that the hypergraph partitioning problem (HGP) is NP-hard~\cite{Lengauer:1990} so that mostly heuristic algorithms are used in practice. 
A successful heuristic to partition large hypergraphs is the \emph{multilevel} approach~\cite{SPPGPOverviewPaper}.
Here, the hypergraph is recursively \emph{contracted} to obtain smaller hypergraphs which should reflect the same basic structure as the input. After applying an \emph{initial partitioning} algorithm to the smallest hypergraph, contraction is undone and, at each level, a
\emph{local search} method is used to improve the partitioning induced by the coarser level. 
The intuition behind this approach is that a good partition at one level of the hierarchy will also be a good partition on the next finer level. Hence, 
depending on the definition of the neighborhood, local search algorithms are able explore local solution spaces very effectively in this setting. 
However, they are also prone to get stuck in local optima~\cite{hMetisKway}. 
The multilevel paradigm helps to some extent, since 
local search has a more global view on the problem on the coarse levels and a very fine-grained view on the fine levels of the multilevel hierarchy. 
In addition, as with many other metaheuristics, multilevel HGP gives better
results if several repeated runs are made with some measures taken to diversify
the search. 

Still even a large number of repeated executions can only scratch the surface of the huge space of possible partitionings. 
In order to explore the global solution space extensively we need more sophisticated metaheuristics. 
This is where memetic algorithms (MAs), \ie, genetic algorithms combined with local search~\cite{KimHKM11}, 
come into play. Memetic algorithms allow for effective exploration (global search) and exploitation (local search) of the solution space.
The general idea behind genetic algorithms is to use mechanisms inspired by biological evolution such as selection, mutation, recombination and survival of the fittest. 
A genetic algorithm (GA) starts with a population of individuals (in our case partitions of the hypergraph) and evolves the population over several generational cycles~(rounds).
In each round, the GA uses a selection rule based on the fitness of the individuals of the population to select good individuals and combines them to obtain improved offspring~\cite{goldbergGA89}. 
When an offspring is generated an eviction rule is used to select a member of the population to be replaced by the new offspring. 
For an evolutionary algorithm it is of major importance to preserve diversity in the population~\cite{baeckEvoAlgPHD96}, i.e., the individuals should not become too similar in order to avoid a premature convergence of the~algorithm.
This is usually achieved by using mutation operations and by using eviction rules that take similarity of individuals into account.

Several genetic and memetic hypergraph partitioning algorithms have already been proposed in the literature~\cite{Areibi00anintegrated,AreibiY04,ArmstrongGAD10,BuiMoon94,Kim2004}. However
\emph{none of them} is considered to be truly competitive with state-of-the-art tools~\cite{Cohoon2003}. We believe that this is due to the fact that all of them employ
 \emph{flat} (i.e., non-multilevel) partitioning algorithms to drive the exploitation of the local solution space.

 Our \emph{main contribution} in this paper therefore is a technique that integrates a memetic algorithm with a \emph{multilevel} hypergraph partitioner.
To this end, we present sophisticated recombination and mutation operators as well as a replacement rule that uses a problem specific similarity measure.
In contrast to previous work~\cite{Areibi00anintegrated,AreibiY04,ArmstrongGAD10,BuiMoon94,Kim2004}, which only considered small and outdated~\cite{MCNCoutdated,ISPD98} ACM/SIGDA benchmark instances~\cite{MCNC} (dating back to the late 1980s), we perform extensive experiments on a large benchmark set containing hypergraphs from several application areas.
Our experiments indicate that our algorithm is able to compute partitions of very high quality and scales well to large networks.
It performs better than KaHyPar, which seems to be the current method of choice among the available hypergraph partitioning tools unless speed is more important than quality~\cite{hs2017sea},
and the state-of-the-art HGP tools hMetis~\cite{hMetisRB,hMetisKway} and PaToH~\cite{PaToH}.
In a setting where competing algorithms get the same fairly large amount of time to compute a solution, 
   our new algorithm computes the best result on $648$ out of the $700$  benchmark instances. 
This is in contrast to previous \emph{non-multilevel} evolutionary algorithms for the problem, which are not considered to be competitive with state-of-the-art tools~\cite{Cohoon2003}.  

\section{Preliminaries}
\subsubsection*{Notation and Definitions.}
An \textit{undirected hypergraph} $H=(V,E,c,$ $\omega)$ is defined as a set of $n$ vertices $V$ and a
set of $m$ hyperedges/nets $E$ with vertex weights $c:V \rightarrow \mathbb{R}_{>0}$ and net 
weights $\omega:E \rightarrow \mathbb{R}_{>0}$, where each net is a subset of the vertex set $V$ (i.e., $e \subseteq V$). The vertices of a net are called \emph{pins}.
We extend $c$ and $\omega$ to sets, i.e., $c(U) :=\sum_{v\in U} c(v)$ and $\omega(F) :=\sum_{e \in F} \omega(e)$.
A vertex $v$ is \textit{incident} to a net $e$ if $v \in e$. $\mathrm{I}(v)$ denotes the set of all incident nets of $v$. 
The set $\Gamma(v) := \{ u~|~\exists~e \in E : \{v,u\} \subseteq e\}$ denotes the neighbors of $v$.
The \textit{size} $|e|$ of a net $e$ is the number of its pins. 
A \emph{$k$-way partition} of a hypergraph $H$ is a partition of its vertex set into $k$ \emph{blocks} $\mathrm{\Pi} = \{V_1, \dots, V_k\}$ 
such that $\bigcup_{i=1}^k V_i = V$, $V_i \neq \emptyset $ for $1 \leq i \leq k$ and $V_i \cap V_j = \emptyset$ for $i \neq j$.
We use $b[v]$ to refer to the block of vertex $v$.
We call a $k$-way partition $\mathrm{\Pi}$ \emph{$\mathrm{\varepsilon}$-balanced} if each block $V_i \in \mathrm{\Pi}$ satisfies the \emph{balance constraint}:
$c(V_i) \leq L_{\max} := (1+\varepsilon)\lceil \frac{c(V)}{k} \rceil$ for some parameter $\mathrm{\varepsilon}$. 
Given a $k$-way partition $\mathrm{\Pi}$, the number of pins of a net $e$ in block $V_i$ is defined as
$\mathrm{\Phi}(e,V_i) := |\{v \in V_i~|~v \in e \}|$. 
For each net $e$, $\mathrm{\Lambda}(e) := \{V_i~|~ \mathrm{\Phi}(e, V_i) > 0\}$ denotes the \emph{connectivity set} of $e$.
The \emph{connectivity} of a net $e$ is the cardinality of its connectivity set: $\mathrm{\lambda}(e) := |\mathrm{\Lambda}(e)|$.
A net is called \emph{cut net} if $\mathrm{\lambda}(e) > 1$.
The \emph{$k$-way hypergraph partitioning problem} is to find an $\varepsilon$-balanced $k$-way partition $\mathrm{\Pi}$ of a hypergraph $H$ that
minimizes an objective function over the cut nets for some $\varepsilon$.
Several objective functions exist in the literature~\cite{Alpert19951,Lengauer:1990}.
The most commonly used cost functions are the \emph{cut-net} metric $\text{cut}(\mathrm{\Pi}) := \sum_{e \in E'} \omega(e)$ and the
\emph{connectivity} metric $(\mathrm{\lambda} - 1)(\mathrm{\Pi}) := \sum_{e\in E'} (\mathrm{\lambda}(e) -1)~\omega(e)$, where $E'$ is the set of all cut nets~\cite{UMPa,donath1988logic}.
Optimizing both objective functions is known to be NP-hard \cite{Lengauer:1990}.
In this paper, we use the connectivity-metric $(\mathrm{\lambda} - 1)(\mathrm{\Pi})$.
\emph{Contracting} a pair of vertices $(u, v)$ means merging $v$ into $u$.
The weight of $u$ becomes $c(u) := c(u) + c(v)$.  We connect $u$ to the former neighbors $\Gamma(v)$ of $v$ by replacing 
$v$ with $u$ in all nets $e \in \mathrm{I}(v) \setminus \mathrm{I}(u)$ and remove $v$ from all nets $e \in \mathrm{I}(u) \cap \mathrm{I}(v)$.
\emph{Uncontracting} a vertex $u$ reverses the contraction.

\vspace*{-.25cm}
\subsection{Related Work}
\label{s:related}
\paragraph{Overview.}Driven by applications in VLSI design and scientific computing, HGP has evolved into a broad research area since the 1990s.
We refer to \cite{Alpert19951,DBLP:conf/dimacs/2012,Papa2007,trifunovic2006parallel} for an extensive overview.
In the following, we focus on issues closely related to the contributions of our paper.
Memetic algorithms~(MAs) were introduced in~\cite{MAs} and formalized in~\cite{RadcliffeS94} as an extension to the concept of genetic algorithms~(GAs)~\cite{Holland:1975}.
While GAs effectively explore the \emph{global} solution space, MAs additionally allow for exploitation of the \emph{local} solution space by incorporating local search
methods into the genetic framework. We refer to \cite{Moscato2010} for an introduction to memetic algorithms.
While several genetic and memetic flat (i.e., non-multilevel) hypergraph partitioning algorithms have been proposed in the literature, \emph{none of them} is considered to be truly competitive with state-of-the-art tools~\cite{Cohoon2003}.
Well-known multilevel HGP software packages with certain distinguishing characteristics include PaToH~\cite{PaToH} (originating from scientific computing),
hMetis~\cite{hMetisRB,hMetisKway} (originating from VLSI design), KaHyPar~\cite{ahss2017alenex,hs2017sea,KaHyPar-R} (general purpose, $n$-level), Mondriaan~\cite{Mondriaan} (sparse matrix
partitioning), MLPart~\cite{MLPart} (circuit partitioning), Zoltan~\cite{Zoltan}, Parkway~\cite{Parkway2.0}, and SHP~\cite{SHP}  (distributed),
UMPa~\cite{DBLP:conf/dimacs/CatalyurekDKU12} (directed hypergraph model, multi-objective), and kPaToH (multiple constraints, fixed vertices)~\cite{Aykanat:2008}.
\vfill\pagebreak
\paragraph{Evolutionary Hypergraph Partitioning.}
Saab and Rao~\cite{SaabR89} present an evolution-based approach for solving a $k$-way multi-objective, multi-constraint hypergraph partitioning problem.
Since the algorithm only works with \emph{one} individual, it does not use any recombination operators. Instead, the solution initially
generated via  bin packing is evolved using a randomized algorithm that moves vertices to different blocks if their
gain is greater than some random value. Hulin~\cite{Hulin1991} provides a GA that uses a coding scheme specifically tailored to circuit bipartitioning along with
crossover and mutation operations that respect the coding.
Bui and Moon~\cite{BuiMoon94} present a steady-state MA for ratio cut bipartitioning of hypergraphs, which uses a weak variation of the FM algorithm~\cite{FMAlgorithm}
as local search engine. To improve the performance of the crossover operation, a preprocessing step re-indexes the vertices by the visiting
order of a weighted depth first search on the clique-representation~\cite{HuMoerder85} of the hypergraph.
Areibi~\cite{Areibi00anintegrated} present a memetic algorithm that combines a GA with a modified version of Sanchis' $k$-way FM algorithm~\cite{HypergraphKFM}.
Areibi and Yang~\cite{AreibiY04} enhance
the MA presented in~\cite{Areibi00anintegrated} with a preprocessing step that clusters and contracts vertices to reduce the complexity of the hypergraphs.
Furthermore, the initial population contains both random as well as good solutions generated using the GRASP heuristic~\cite{FeoRS94}.
Armstrong et al.~\cite{ArmstrongGAD10} propose a $k$-way MA that performs crossover, mutation and local search on multiple individuals in parallel.
The traditional FM algorithm~\cite{FMAlgorithm} and Sanchis' $k$-way FM version~\cite{HypergraphKFM} are used for local search.
Kim et al.~\cite{Kim2004} present a steady-state MA for hypergraph bipartitioning, which uses a modified FM algorithm that works with lock-gains~\cite{KimM04}.
Note that \emph{none} of these algorithms makes use of the multilevel~paradigm.

\paragraph{Evolutionary Graph Partitioning.}
We refer to the survey of Kim \etal \cite{KimHKM11} for a general overview and more material on genetic approaches for graph partitioning.
Soper et al. \cite{soper2004combined} provide the first algorithm that combined an evolutionary search algorithm within a multilevel graph partitioner. Here, crossover and mutation operators compute edge biases based on the input individuals. 
A similar approach based on perturbations of edge weights has been used by Delling \etal\cite{delling2010graph}.
Benlic et al. \cite{benlichao2010} provide a multilevel memetic algorithm for balanced graph partitioning. 
PROBE \cite{ChardaireBM07} is a metaheuristic which can be viewed as a genetic algorithm without selection. It outperforms other metaheuristics, but it is restricted to the case $k=2$ and $\varepsilon=0$.
KaHIP~\cite{sandersschulz2013} contains  KaFFPaE~\cite{kaffpaE}, which has a general recombine operator framework based on a multilevel algorithm.

\subsection{$k$-way Hypergraph Partitioning using KaHyPar}
Since our memetic algorithm builds on top of the KaHyPar framework, we briefly review its core components.
While traditional multilevel HGP algorithms contract matchings or clusterings and therefore
work with a coarsening hierarchy of $\Oh{\log n}$ levels, KaHyPar instantiates the multilevel paradigm in the
extreme $n$-level version, removing only a \emph{single} vertex between two levels.
Vertex pairs $(u,v)$ to be contracted are determined using the heavy-edge rating function $r(u,v) := \sum_{e \in E'}  \omega(e)/(|e| - 1)$, where $E' := \{\mathrm{I}(u) \cap \mathrm{I}(v)\}$.
The coarsening process stops
as soon as the number of vertices drops below a certain threshold or no more contractions are possible. The framework currently contains
two coarsening algorithms.
The first algorithm~\cite{KaHyPar-R} contracts vertices in decreasing rating score order using a priority queue to store and update the ratings. 
 The second algorithm~\cite{ahss2017alenex} immediately contracts each vertex with its highest-rated neighbor in \emph{random} order.
After coarsening, a portfolio of simple algorithms is used to create an initial partition of the coarsest hypergraph. During uncoarsening,
strong localized local search heuristics based on the FM algorithm~\cite{FMAlgorithm,HypergraphKFM} are used to refine
the solution by moving vertices to other blocks in the order of improvements in the optimization objective.
The framework provides a recursive bisection
algorithm to optimize the cut-net metric (KaHyPar-R~\cite{KaHyPar-R}) as well as a direct $k$-way algorithm to optimize the $(\lambda-1)$ metric (KaHyPar-K~\cite{ahss2017alenex}).
Recently,
Heuer and Schlag~\cite{hs2017sea} integrated an improved coarsening scheme into KaHyPar-K that incorporates global information
about the structure of the hypergraph into the coarsening process. It uses community detection in a preprocessing step and prevents
inter-community contractions during coarsening. This version is referred
to as KaHyPar-CA. Unless mentioned otherwise, we use the default
configurations provided by~the~authors\footnote{\url{https://github.com/SebastianSchlag/kahypar/tree/master/config}}. 

\vspace*{-.35cm}
\section{Memetic Multilevel Hypergraph Partitioning}
\vspace*{-.25cm}
We now explain the components of our memetic multilevel hypergraph partitioning algorithm.
Given a hypergraph $H$ and a time limit $t$, the algorithm starts by creating an initial
population of $\mathcal{P}$ \emph{individuals}, which in our case correspond to $\varepsilon$-balanced $k$-way partitions of $H$.
The population size $|\mathcal{P}|$ is determined dynamically by first measuring the time $t_\text{I}$ spend to create one individual.
Then $\mathcal{P}$ is chosen such that the time to create $|\mathcal{P}|$ individuals is a certain percentage $\delta$ of the total running time $t$:
$|\mathcal{P}| := \max(3,\min(50,\delta\cdot(t/t_I)))$, where $\delta$ is a tuning parameter.
The lower bound on the population size is chosen to ensure a certain minimum of diversity, while
the upper bound is used to ensure convergence.
In contrast to previous approaches~\cite{Areibi00anintegrated,ArmstrongGAD10,BuiMoon94,Hulin1991,Kim2004} the population is not
filled with randomly generated individuals, but \emph{high-quality} solutions computed by KaHyPar-CA.

To judge the \emph{fitness} of an individual
we use the connectivity $(\lambda -1)(\mathrm{\Pi})$ of its partition $\mathrm{\Pi}$.
The initial population is evolved over several generational cycles using the \emph{steady-state} paradigm~\cite{EvoComp}, i.e., we generate
only \emph{one} offspring per generation.
Our two-point and multi-point recombination operators described in Section~\ref{sec:combine} improve the average quality of the population
by effectively combining different solutions to the HGP problem.
In order to sufficiently explore the global search space and to prevent premature convergence, it is important to keep
the population diverse~\cite{baeckEvoAlgPHD96}. This becomes even more relevant in our case, since with KaHyPar-CA we use powerful
heuristics to exploit the local solution space.
Previous work on evolutionary algorithms for HGP~\cite{AreibiY04,ArmstrongGAD10,BuiMoon94,Hulin1991,Kim2004} used simple mutations that
change the block of each vertex uniformly at random with a small probability. In contrast to these simple, problem agnostic operators,
we propose  mutation operators based on V-cycles~\cite{WalshawVcycle} that
exploit knowledge of the problem domain and create offspring solutions in the \emph{vicinity} of
the current population.
Furthermore in Section~\ref{sec:replace} we propose a replacement strategy which considers
fitness \emph{and} similarity to determine the individual to be evicted from the population.

\subsection{Recombination Operators}\label{sec:combine}
\begin{figure}[t!] 
    \vspace*{-.25cm}
  \centering
  \includegraphics[width=.8\textwidth]{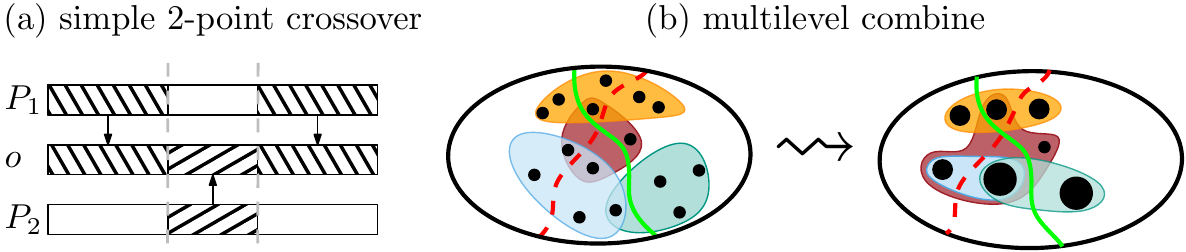}
  \caption{(a) Traditional, problem agnostic crossover operation to combine parent partitions $P_1$ and $P_2$ to offspring $o$.
    (b) Recombination using modified multilevel coarsening to combine two partitions (dashed red line and solid green line).
    Each cut net $e$ remains in the coarse hypergraph and maintains its connectivity $\lambda(e)$ regarding both partitions.}\label{fig:combine}
    \vspace*{-.5cm}
\end{figure}

\vspace*{-.125cm}
The evolutionary algorithms for HGP presented in Section~\ref{s:related} 
use simple multi-point crossover operators which split the parent partitions into several parts and then combine these parts to form new offspring (see Figure~\ref{fig:combine} (a)).
Since these operators do not take the structure of the hypergraph into account, offspring solutions may have
considerably worse fitness than their parents. By generalizing the recombine operator framework presented in \cite{kaffpaE} from
graphs to hypergraphs, our two-point recombine operators described in this section assure that the fitness of the offspring is
\emph{at least as good as the best of both parents}. The edge frequency based multi-point recombination operator described afterwards
gives up this property, but still generates good offspring.

\paragraph{Two-Point Recombine.}
The operator starts with selecting parents for recombination using binary tournament selection (without replacement)~\cite{BlickleT96}.
Two individuals $I_1$ and $I_2$ are chosen uniformly at random from $\mathcal{P}$ and
the individual with better fitness (i.e., lower $(\lambda-1)$ objective) becomes the first parent~$P_1$. This process is then repeated
to determine the second parent $P_2$. A tournament size of two is chosen to keep the selection
pressure low and to avoid premature convergence, since all our individuals already constitute high-quality solutions.
Both individuals/partitions are then used as input of a modified multilevel partitioning scheme as follows:
During coarsening, two vertices $u$ and $v$ are only allowed to be contracted if \emph{both parents agree
on the block assignment of both vertices}, i.e., if $b_1[u] = b_1[v] \wedge b_2[u] = b_2[v]$. This
is a generalization from multilevel evolutionary GP, \ie\cite{kaffpaE}, where \emph{edges running between two blocks are not eligible} for contraction
and therefore remain in the graph. In other words, our generalization allows two vertices of the same cut net to be contracted as long as the input individuals agree that they belong to the same block.
For HGP, this restriction ensures that cut nets $e$ remain in the coarsened hypergraph
\emph{and} maintain their connectivity $\lambda(e)$ regarding \emph{both} partitions. This modification is important for our optimization objective,
because it allows us to use the partition of the \emph{better} parent as initial partition of the offspring.
Since we can skip the initial partitioning phase and therefore do not need a sufficiently large number of vertices in the coarsest hypergraph to compute a good initial partition~\cite{hMetisKway}, we alter the stopping criterion of the coarsening phase such that it stops when no more contractions are possible.
The high quality solution of the coarsest hypergraph contains two different classes of vertices: Those for which both parent partitions agree on a block assignment
and those for which they don't (see Figure~\ref{fig:combine} (b) for an example).
During the uncoarsening phase, local search algorithms can then use this initial partitioning to (i) exchange good parts of the solution on the coarse
levels by moving few vertices and (ii) to find the best block assignment for those vertices, for which the parent partitions disagreed.
Since KaHyPar's refinement algorithms guarantee nondecreasing solution quality, the fitness of offspring solutions generated using
this kind of recombination is always \emph{at least as good as the better of both parents}.

\paragraph{Edge-Frequency Multi-Recombine.}
The operator described previously is restricted to recombine $p=2$ partitions to improved offspring of nondecreasing quality.
Sanders and Schulz~\cite{kaffpaE} specifically restrict their operators to this case, arguing that in the course of the algorithm
a series of two-point recombine operations to some extend emulates a multi-point recombine.
We here present a reasonable multi-point recombine operation to partially evaluate this hypothesis in our experimental evaluation.
Our recombine operator uses the concept of (hyper)edge frequency~\cite{Wichlund98} to pass information about the cut nets of the $t$ \emph{best individuals} in the population on to new offspring.
The frequency $f(e)$ of a net $e$ hereby refers to the number of times it appears in the cut in the $t$ best solutions: $f(e) := |\{I \in t~|~\lambda(e) > 1\}|$. We use $t = \lceil \sqrt{|\mathcal{P}|} \rceil$,
which is a common value in evolutionary algorithms~\cite{delling2010graph}.
Our multi-recombine operator then uses this information to create a \emph{new} individual in the following way. The
coarsening algorithm is modified to prefer to contract vertex pairs $(u,v)$ which share a large number of small, low-frequency nets. This is achieved
by replacing the standard heavy-edge rating function of KaHyPar with the rating function~\cite{Wichlund98} shown in Eq.~\ref{ef}:

\begin{equation}\label{ef}
r(u,v) := \frac{1}{c(v) \cdot c(u)}~\sum \limits_{e \in \{\mathrm{I}(v) \cap \mathrm{I}(u)\}}  \frac{\exp(-\gamma f(e))}{|e|}.
\end{equation}

This rating function disfavors the contraction of vertex pairs incident to cut nets with high frequency, because these nets 
are likely to appear in the cut of high quality solutions. The tuning parameter $\gamma$ is used as a damping factor.
After coarsening stops, we compute an initial partition of the coarsest hypergraph using KaHyPar's initial partitioning algorithms and refine
it during the uncoarsening and local search phase.

\subsection{Mutation Operations and Diversification}\label{sec:mutate}
We define two mutation operators based on V-cycles.
All operators are applied to a random individual $I$ of the current~population. 
The main idea of V-cycle based mutation operators is to reuse an already computed partition as input for the multilevel approach and to iterate
coarsening and local search phases several times using different seeds for randomization. This approach has been applied successfully
in evolutionary GP~\cite{kaffpaE}, therefore we also adopt it for HGP.
Similar to the recombine operator described in Section~\ref{sec:combine}, the quality of the solution is maintained by only contracting vertex pairs $(u,v)$
belonging to the same block ($b[u] = b[v]$). By distinguishing two possibilities for initial partitioning, we define two different
mutation operators: The first one uses the current partition of the individual as initial partition of the coarsest hypergraph and guarantees nondecreasing
solution quality. The second one employs KaHyPar's portfolio of initial partitioning algorithms to compute a new solution for the coarsest hypergraph.
During uncoarsening, local search algorithms improve the solution quality and thereby further mutate the individual. 
Since the second operator computes a new initial partition which might be different from the original partition of $I$, the fitness of offspring generated by this operator can be worse than the fitness of $I$.

\subsection{Replacement Strategy}\label{sec:replace}
All recombination and mutation operators create one new offspring $o$.
In order to keep the population diverse, we evict the individual \emph{most similar} to the offspring among all individuals whose fitness is equal to or worse than $o$.
Previous work on bipartitioning~\cite{BuiMoon94,Kim2004} used the Hamming distance as a metric to measure the similarity between partitions.
We propose a more sophisticated similarity measure that takes into account the connectivity $\lambda(e)$ of each cut net $e$.
For each individual, we compute the multi-set $D :=  \{(e, m(e)) : e \in E \} $, where $m(e) := \lambda(e) - 1$ is the
multiplicity (\ie number of occurrences) of $e$. Thus each cut net $e$ is represented $\lambda(e) - 1$ times in $D$.
The difference of two individuals $I_1$ and $I_2$ is the computed as $d(I_1, I_2) := | D_1 \ominus D_2 |$, where  $\ominus$ is the symmetric difference.

\section{Experimental Evaluation}\label{sec:eval}
\subsubsection*{System and Methodology.}\label{Methodology}
We implemented the memetic algorithm described in the previous section using the latest version of the KaHyPar framework.
The code is written in C++ and compiled using g++-5.2 with flags \texttt{-O3} \texttt{-mtune=native} \texttt{-march=native}.
We refer to the algorithm presented in this paper as EvoHGP.
All experiments are performed on a cluster with $512$ nodes, where each node has two Intel Xeon E5-2670 Octa-Core (Sandy Bridge) processors 
clocked at $2.6$ GHz, $64$~GB main memory, $20$ MB L3- and 8x256 KB L2-Cache and runs RHEL 7.4.

We compare EvoHGP with two different configurations of KaHy-Par-CA~\cite{hs2017sea},
as well as to the $k$-way (hMetis-K) and the recursive bisection variant (hMetis-R) of hMetis 2.0 (p1)~\cite{hMetisRB,hMetisKway}, and to PaToH 3.2~\cite{PaToH}. 
These HGP libraries were chosen because they provide the best solution quality~\cite{hs2017sea,ahss2017alenex}.
The first configuration of KaHyPar-CA  corresponds to the default configuration as described in \cite{hs2017sea}. 
Since it is known that global search strategies are more effective than plain restarts~\cite{KahipWFCycles}, we
augment KaHyPar-CA with V-cycles (in a similar fashion as the first mutation operator) using
a maximum number of $100$ V-cycle iterations per partitioner call. This \emph{new} enhanced version of KaHyPar-CA
constitutes the second configuration and is referred to as KaHyPar-CA-V.
hMetis and PaToH are configured as described in \cite{hs2017sea}.
Since PaToH ignores the random seed if configured to use the quality preset, we only report result of the default configuration (PaToH-D). 
For all five algorithms we perform repeated runs using different random seeds for each run. 

To evaluate the impact of different algorithmic components of our algorithm in Section~\ref{subs:config} each EvoHGP configuration gets \emph{two} hours time
\emph{per} test instance to compute a solution. For the final evaluation in Section~\ref{subsec:eval} all algorithms get \emph{eight} hours time per test instance.
In both cases, we perform five repetitions with different seeds for each test instance
and algorithm. Due to the large amount of computing time necessary to perform these experiments, we always partition $16$ instances in
parallel on a single node.
We use the \emph{arithmetic mean} when averaging over solutions of the same instance and the \emph{geometric mean} when averaging over different instances
in order to give every instance a comparable influence on the final result.
In order to compare EvoHGP with the different algorithms, we present two kinds of plots:
\emph{Convergence plots}~\cite{kaffpaE} show the evolution of solution quality over time normalized by instance size, while \emph{performance plots}~\cite{KaHyPar-R} are used to compare
the best solutions of all algorithms on a per-instance basis. 

\paragraph{Convergence Plots.}
We start by explaining how to compute the data for a single instance $I$, i.e., a $k$-way partition of a hypergraph $H$.
Whenever an algorithm computes a partition that improves the solution quality, it
reports a pair ($t$, $(\lambda - 1)$), where the timestamp $t$ is the currently elapsed time.
For $r$ repetitions with different seeds $s$, these $r$ sequences $T_s^I$ of
pairs are merged into one sequence  $T^I$ of triples $(t, s, (\lambda - 1))$, which is sorted by the timestamp $t$.
Since we are interested in the \emph{evolution} of the solution quality, we compute the sequence $T_{\text{min}}^I$ representing \emph{event-based} average values.
We start by computing the average connectivity $\overline{c}$ and the average time $\overline{t}$ using the first pair ($t$, $(\lambda - 1)$) of all $r$ sequences $T_s^I$ and insert
$(\overline{t}, \overline{c})$ into $T_{\text{min}}^I$. We then sweep through the remaining entries $(t, s, (\lambda - 1))$ of $T^I$. Each entry corresponds
to a partition computed at timestamp $t$ using seed $s$ that improved the solution quality to $(\lambda - 1)$. For each entry we therefore replace the old connectivity value of seed $s$
that took part in the computation of $\overline{c}$ with the new value $(\lambda - 1)$, recompute $\overline{c}$ and insert a new pair $(t, \overline{c})$ into $T_{\text{min}}^I$.
$T_{\text{min}}^I$ therefore represents the evolution of the average solution quality $\overline{c}$ for instance $I$ over time.
In a final step, we create the \emph{normalized} sequence $N_{\text{min}}^I$, where each entry $(t, \overline{c})$ in $T_{\text{min}}^I$ is replaced by
$(t_n, \overline{c})$ where $t_n := t/t_I$ and $t_I$ is the average time that KaHyPar-CA needs to compute a $k$-way partition of $H$.
Average values over \textit{multiple instances} are then obtained as follows:
All sequences $N^I_{\text{min}}$ of pairs $(t_n,\overline{c})$ are merged into a sequence $N_{\text{min}}$ of triples  $(t_n, \overline{c}, I)$,
which is then sorted by $t_n$. The final sequence $S_{\mathcal{G}}$ presenting event-based \emph{geometric} averages values is then computed as follows:
We start by computing the average normalized time $\overline{t_n}$ and the geometric mean connectivity $\mathcal{G}$ over all instances $I$ using the first value of all $N^I_\text{min}$ and
insert $(\overline{t_n}, \mathcal{G})$ into $S_{\mathcal{G}}$. We then sweep through the remaining entries of $N_{\text{min}}$.
For each entry $(t_n, \overline{c}, I)$, we replace the old connectivity value of $I$ that took part in the computation of $\mathcal{G}$ with the new value $\overline{c}$,
recompute $\mathcal{G}$ and insert $(t_n, \mathcal{G})$ into $S_{\mathcal{G}}$. The sequence $S_{\mathcal{G}}$ therefore represents the evolution of the solution quality
averaged over all instances~and~repetitions.

\paragraph{Performance Plots.}
These plots relate the smallest minimum connectivity of all algorithms to the corresponding connectivity produced by each algorithm on a per-instance basis.
For each algorithm, these ratios are sorted in increasing order. The plots use a cube root scale for both axes to reduce right skewness~\cite{st0223}
and show $1-(\text{best}/\text{algorithm})$ on the y-axis to highlight the instances were each partitioner performs badly.
A point close to one indicates that the partition produced by the corresponding algorithm was considerably worse than the
partition produced by the best algorithm. A value of zero therefore indicates that the corresponding algorithm produced the best solution.
Points above one correspond to infeasible solutions that violated the balance constraint.
Thus an algorithm is considered to outperform another algorithm if its corresponding ratio values are below those of the other algorithm.
In order to include instances with a cut of zero into the results, we set the corresponding cut values to \emph{one} for ratio computations.

\vspace*{-.5cm}
\subsubsection*{Benchmark Instances.} \label{Instances}
We evaluate our algorithm on a representative subset of $100$ hypergraphs from the benchmark set of Heuer and Schlag~\cite{hs2017sea}\footnote{The benchmark set was downloaded from \url{http://algo2.iti.kit.edu/schlag/sea2017/}.}, which
contains instances from four benchmark sets: the ISPD98 VLSI Circuit Benchmark Suite~\cite{ISPD98},
the DAC 2012 Routability-Driven Placement Contest~\cite{DAC2012},
the University of Florida Sparse Matrix Collection~\cite{FloridaSPM}, and the international SAT Competition 2014~\cite{SAT14Competition}.
Sparse matrices are translated into hypergraphs using the row-net model~\cite{PaToH}, i.e., each row is treated as a net and each column as a vertex.
SAT instances are converted to three different representations: For literal hypergraphs, each boolean \emph{literal} is mapped to one vertex and each clause constitutes a net~\cite{Papa2007}, while in the \emph{primal} model each variable is represented by a vertex and each clause is represented by a net.
In the \emph{dual} model the opposite is the case~\cite{FormulaPartitioning14}.
The latter two models are more common in the SAT solving community than the literal model proposed in~\cite{Papa2007}.
All hypergraphs have unit vertex and net weights. 
An overview of our benchmark sets is given in Tables~\ref{tbl:instancessmall} and \ref{tbl:instanceslarge} in Appendix~\ref{app:hypergraphs}.
To compare EvoHGP with the best competing algorithms, all $100$ hypergraphs are partitioned into $k \in \{2,4,8,16,32,64,128\}$ blocks with $\varepsilon = 0.03$.
For each hypergraph $H$ and each value of $k$, a $k$-way partition $H$ is considered to be \emph{one} test instance, resulting in a total of $700$ instances.
In order to save running time, we choose a subset of 25 hypergraphs shown in Table~\ref{tbl:instancessmall}, $k=32$, and $\varepsilon = 0.03$ to evaluate the impact of different algorithmic components of our algorithm
 (recombine and mutation operations) before we run the algorithms on the large benchmark set.

\subsection{Influence of Algorithmic Components}\label{subs:config}
All configurations determine their population size $\mathcal{P}$ dynamically such that $\delta=15\%$ of the total time is spent to create the initial population.
According to the results of Wichlund and Aas~\cite{Wichlund98},
\begin{wrapfigure}{r}{8.5cm}
\centering
\vspace*{-.5cm}
\begin{knitrout}
\definecolor{shadecolor}{rgb}{0.969, 0.969, 0.969}\color{fgcolor}

{\centering \includegraphics[width=.5\textwidth]{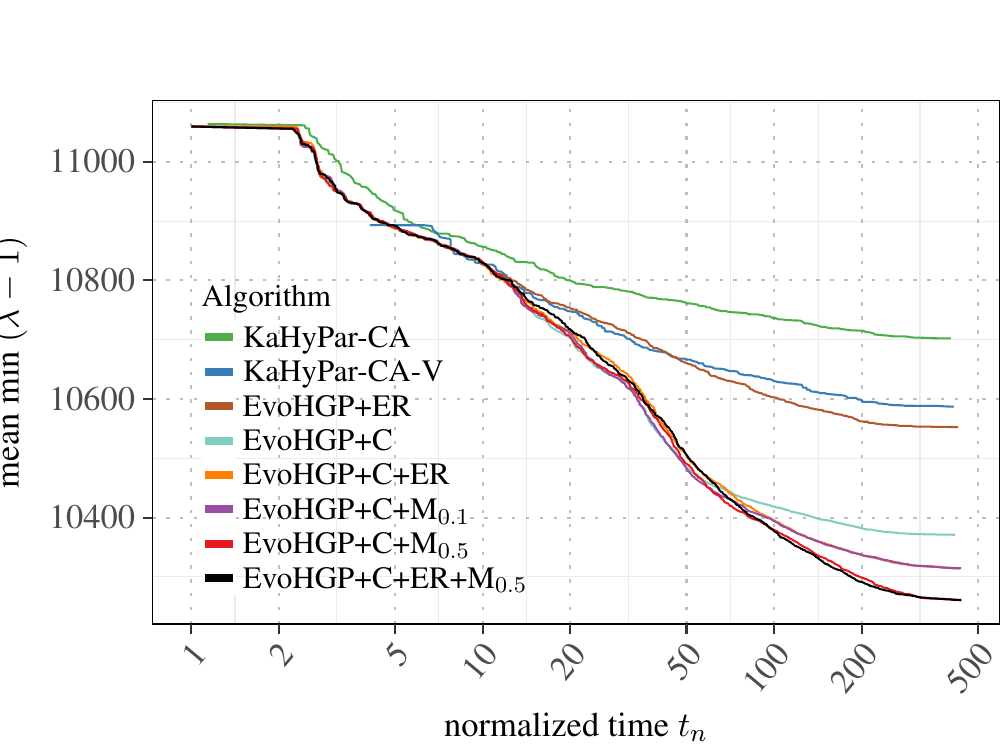} 

}
\end{knitrout}
\caption{Influence of EvoHGPs algorithmic components.}\label{fig:config}
\vspace*{-.75cm}
\end{wrapfigure}
the damping factor $\gamma$ used for edge frequency calculations is set to $\gamma=0.5$.
We use a naming scheme to refer to different configurations of our algorithm. All configuration names start with EvoHGP followed by abbreviations 
for the added recombine and mutation operations (multiple abbreviations are used to add multiple operations).
Abbreviation +C refers to EvoHGP using two-point recombine operations, +ER refers to EvoHGP using multi-recombine operations,
and finally +M$_{x}$ adds mutation operations with a mutation chance of $x$.
Whenever a mutation operation is performed, both operators have a 50 percent change of being chosen.
Figure~\ref{fig:config} compares different configurations of EvoHGP. 
Of all configurations, EvoHGP+ER, which relies only on multi-point recombine operations,
performs worst, being only slighthly better than KaHyPar-CA-V.
Comparing its performance with EvoHGP+C (which uses only two-point recombine operations), we can see that it is indeed beneficial to guarantee nondecreasing solution quality for
combine operations. However combining both recombination operators results in a performance similar to EvoHGP+C+M$_{0.1}$. This can be explained by the fact that
multi-recombines also act as mutation operator in that they don't guarantee nondecreasing performance.
Due to the fact that the strong multilevel local search engine KaHyPar-CA computes high quality solutions, we see that a significant amount of mutations
is necessary to ensure diversity in the population. While EvoHGP+C+M$_{0.1}$ ($10\%$ mutation chance performed best for evolutionary graph partitioning in~\cite{kaffpaE}) performs
equally well as EvoHGP+C+ER, increasing the mutation rate to $50\%$ (EvoHGP+C+M$_{0.5}$) improves the overall performance of the algorithm. Moreover, we see that using both recombination operators
and mutations (EvoHGP+C+ER+M$_{0.5}$) also performs well.
Since EvoHGP+C+M$_{0.5}$ and EvoHGP+C+ER+M$_{0.5}$ show the best convergence behavior, we restrict ourselves to these configurations for the remaining experiments performed in the paper.

\vspace*{-.25cm}
\subsection{Evaluation}\label{subsec:eval}
\vspace*{-.25cm}
\begin{figure}[t!]
\centering
\begin{knitrout}
\definecolor{shadecolor}{rgb}{0.969, 0.969, 0.969}\color{fgcolor}

{\centering \includegraphics[width=\textwidth]{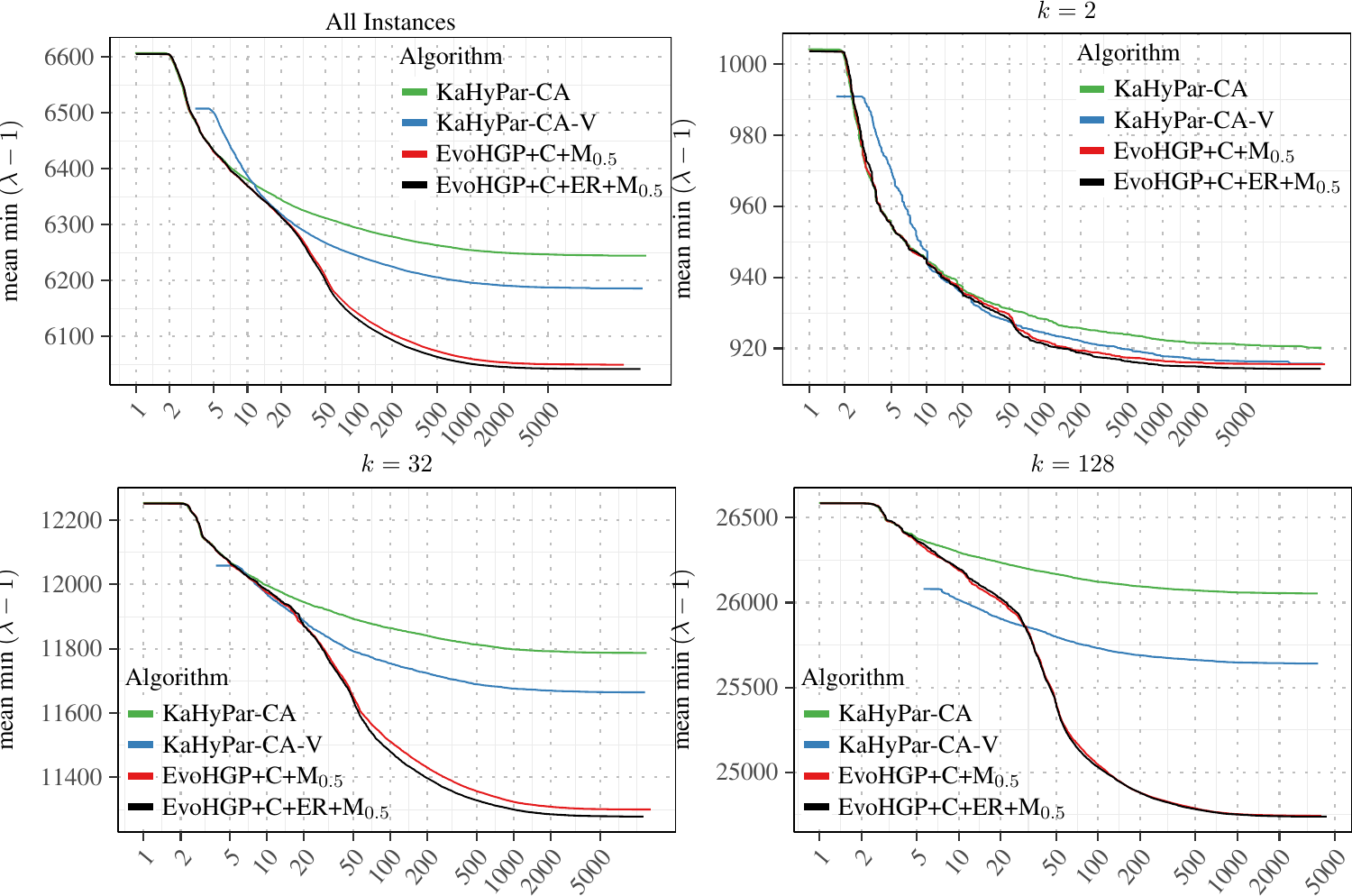}

}
\caption{Convergence plots for all instances and for different values of $k$.}\label{fig:convergence}
\end{knitrout}
\vspace*{-.5cm}
\end{figure}

\begin{table}
  \centering
\caption{Average improvement in connectivity metric over both KaHyPar-CA and the new version KaHyPar-CA-V.}~\label{tbl:imp}
\begin{tabular}{@{\extracolsep{2pt}}rcccc@{}}
  \toprule
  & \multicolumn{2}{c}{KaHyPar-CA vs. EvoHGP} & \multicolumn{2}{c}{KaHyPar-CA-V vs. EvoHGP}                 \\
  \cline{2-3}\cline{4-5}
  $k$    & +C+M$_{0.5}$                    & +C+ER+M$_{0.5}$ & +C+M$_{0.5}$ & +C+ER+M$_{0.5}$\\
  \midrule
  all & 3.3\%                           & 3.4\%           & 2.3\%        & 2.4\%           \\
2     & 0.9\%                           & 0.9\%           & 0.3\%        & 0.4\%           \\
4     & 1.3\%                           & 1.4\%           & 0.8\%        & 1.0\%           \\
8     & 2.7\%                           & 2.9\%           & 1.9\%        & 2.0\%           \\
16    & 3.5\%                           & 3.6\%           & 2.5\%        & 2.6\%           \\
32    & 4.3\%                           & 4.6\%           & 3.2\%        & 3.5\%           \\
64    & 4.9\%                           & 5.0\%           & 3.5\%        & 3.6\%           \\
128   & 5.4\%                           & 5.4\%           & 3.7\%        & 3.7\%           \\
\bottomrule
\end{tabular}
\end{table}

\begin{figure}[t]
  \centering
   \includegraphics[width=.45\textwidth]{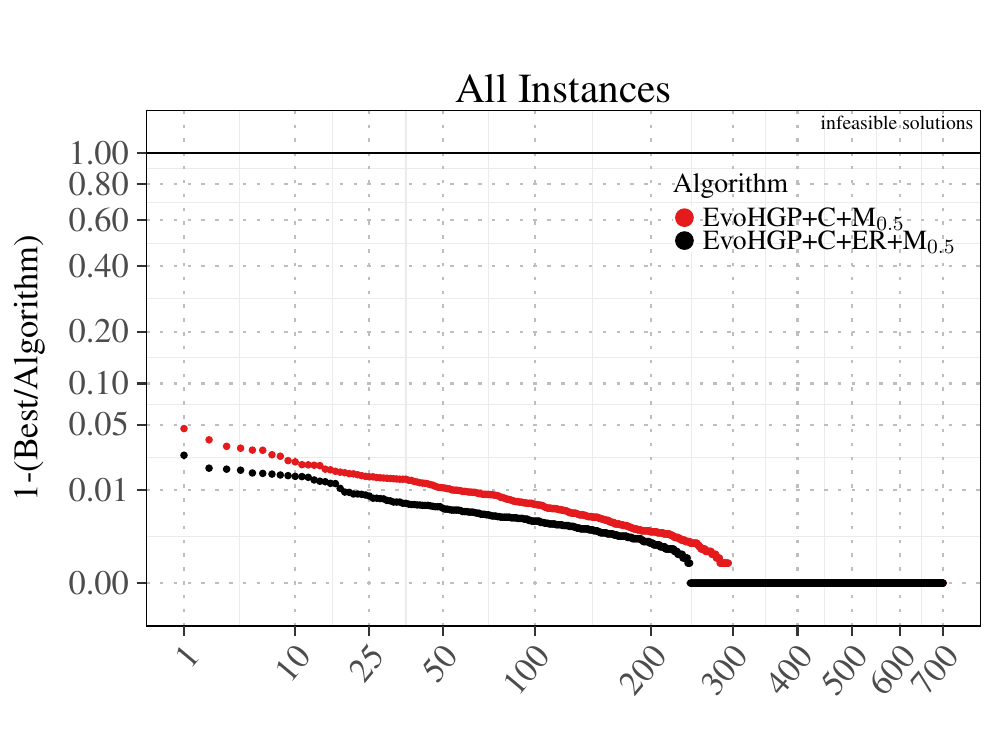} 
\includegraphics[width=.45\textwidth]{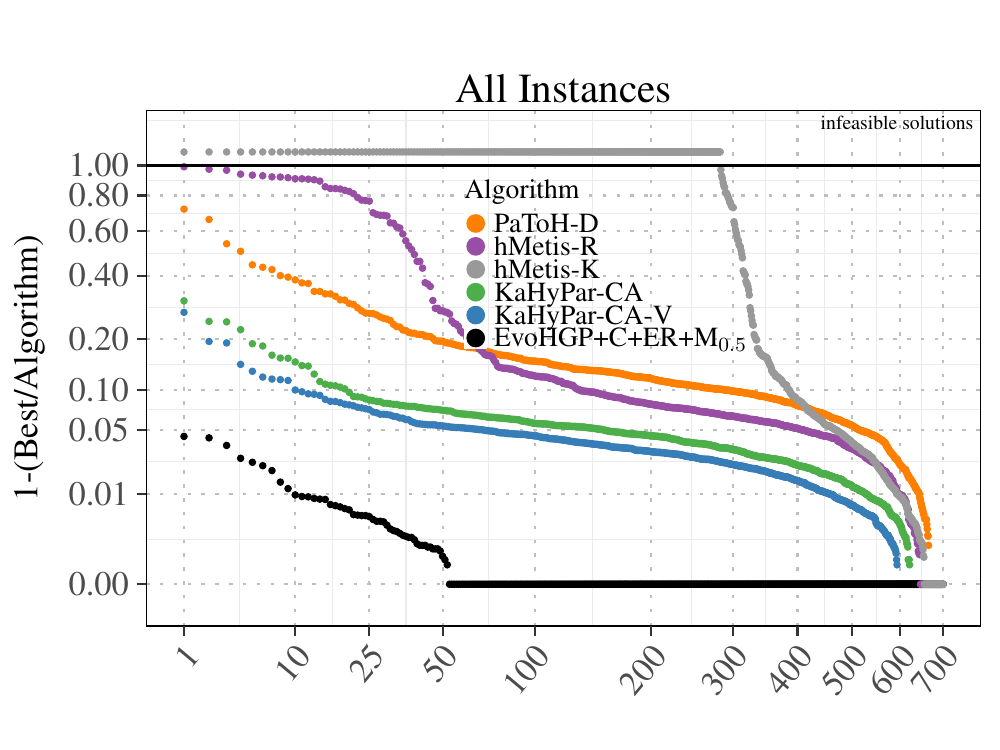}
\caption{Performance plots comparing the two final configurations of our evolutionary algorithm (left) and the best configuration of EvoHGP to the strongest non-evolutionary algorithms (right).
EvoHGP+C+ER+M$_{0.5}$ outperforms the non-evolutionary algorithms on almost all instances.}~\label{fig:performance}
\end{figure}

We now switch to our large benchmark set to evaluate the performance of the different algorithms under consideration.
Table~\ref{tbl:imp} and Figure~\ref{fig:convergence} compare the performance of our memetic algorithms with repeated executions of
KaHyPar-CA and KaHyPar-CA-V. When looking at convergence plots, note that KaHyPar-CA-V starts later than all other algorithms and has an initially better solution quality.
This is due to the fact it uses up to $100$ V-cycles before reporting the first solution.
The improvements of our memetic algorithms increase with increasing $k$. This is expected as the search space of possible partitionings
increases with the number of blocks. 
Looking at Table~\ref{tbl:imp}, we see that both memetic algorithms on average outperform the best partitioner currently available (KaHyPar-CA), culminating in an improvement of $5.4\%$
for $k=128$. Furthermore both EvoHGP+C+M$_{0.5}$ and EvoHGP+C+ER+M$_{0.5}$ are able to improve upon the \emph{new} V-cycling version KaHyPar-CA-V for all values of $k$ and performs
$3\%$ better on average than KaHyPar-CA-V for $k \geq 32$. While the difference in solution quality between both memetic algorithms is small on average,
a Wilcoxon matched pairs signed rank test~\cite{wilcoxon} (using a 1\% significance level) reveals that the improved solution quality of EvoHGP+C+ER+M$_{0.5}$
is statistically significant ($Z=-2.992857$, $p=0.002763795$). This is also confirmed by the performance plot shown in Figure~\ref{fig:performance} (left).

Looking at the performance plot that compares the strongest non-evolutionary algorithms with the strongest memetic configuration in Figure~\ref{fig:performance} (right),
we see that EvoHGP+C+ER+M$_{0.5}$ performs significantly better than \emph{all} other algorithms. It produces the best partitions for $648$ of the $700$ instances.
It is followed by KaHyPar-CA-V (107), KaHyPar-CA (79), hMetis-R (55), hMetis-K (46), and PaToH-D (35).
Note that for some instances, multiple partitioners computed the same solution.
Comparing the best solutions of EvoHGP+C+ER+M$_{0.5}$ to each partitioner individually, it produced better partitions than PaToH-D, hMetis-K, hMetis-R, KaHyPar-CA,
KaHyPar-CA-V in $661$, $644$, $636$, $609$, $585$ cases, respectively.

This shows that even a large number of repeated executions helps only partially to explore the huge space of possible partitionings.
By combining effective exploration (global search) with exploitation (in our case using powerful $n$-level HGP algorithms)
our memetic algorithm can effectively help to break out of local minima and hence explore the global solution space more extensively.

\vspace*{-.25cm}
\section{Conclusion and Future Work}
\vspace*{-.25cm}
EvoHGP is the first \emph{multilevel} memetic algorithm to tackle the balanced hypergraph partitioning problem. 
Key components of our contribution are new effective multilevel recombine and mutation operations that incorporate information about the best solutions in the coarsening process
and provide a large amount of diversity.
Experiments comparing EvoHGP with a V-cycling version of KaHyPar-CA and the well known HGP tools hMetis and PaToH indicate that our evolutionary algorithm computes by far the best solutions on almost all instances.
This confirms our conjecture that previous attempts to solve the HGP problem using memetic algorithms failed to be competitive with
state-of-the-art tools because (i) only flat partitioning algorithms were used to drive the exploitation phase and (ii) evolutionary
operators were problem agnostic and thus did not incorporate enough structural information into the algorithm.
We therefore believe that EvoHGP is helpful in a wide area of application areas in which solution quality is of major importance.
In the future, it would be interesting to apply EvoHGP in such application areas and to try other domain specific recombine operators that offer more specific knowledge of the application domain.
In addition, it may be worth to investigate shared-memory parallelization as in~\cite{ArmstrongGAD10} or a distributed memory parallelization based on islands as in~\cite{kaffpaE}.
Lastly, we plan to integrate our algorithm in the KaHyPar framework.

\vfill\pagebreak
\bibliography{p50-schlag} 
\vfill\pagebreak

\begin{appendix}
\vfill\pagebreak
\section{Benchmark Hypergraphs}~\label{app:hypergraphs}
\begin{table}
        \caption{Basic properties of our small benchmark subset.}
\label{tbl:instancessmall}

\centering
\begin{tabular}{@{\extracolsep{2pt}}lrrrlrrr@{}}
  \toprule
Hypergraph                      & $n$     & $m$    & $p$     & Hypergraph               & $n$    & $m$    & $p$     \\
\cline{5-8}
\cline{1-4}
\multicolumn{4}{c}{ISPD98}    & \multicolumn{4}{c}{SAT14Primal}                                                   \\
\cline{5-8}
\cline{5-8}
\cline{1-4}
\cline{1-4}
ibm06                           & 32498   & 34826  & 128182  & 6s153                    & 85646  & 245440 & 572692  \\
ibm07                           & 45926   & 48117  & 175639  & aaai10-planning-ipc5     & 53919  & 308235 & 690466  \\
ibm08                           & 51309   & 50513  & 204890  & atco\_enc2\_opt1\_05\_21 & 56533  & 526872 & 2097393 \\
ibm09                           & 53395   & 60902  & 222088  & dated-10-11-u            & 141860 & 629461 & 1429872 \\
ibm10                           & 69429   & 75196  & 297567  & hwmcc10-timeframe        & 163622 & 488120 & 1138944 \\
\cline{1-4}
\cline{5-8}
\multicolumn{4}{c}{SAT14Dual} & \multicolumn{4}{c}{SPM}                                                           \\
\cline{5-8}
\cline{1-4}
6s133                           & 140968  & 48215  & 328924  & laminar\_duct3D          & 67173  & 67173  & 3833077 \\
6s153                           & 245440  & 85646  & 572692  & mixtank\_new             & 29957  & 29957  & 1995041 \\
6s9                             & 100384  & 34317  & 234228  & mult\_dcop\_01           & 25187  & 25187  & 193276  \\
dated-10-11-u                   & 629461  & 141860 & 1429872 & RFdevice                 & 74104  & 74104  & 365580  \\
dated-10-17-u                   & 1070757 & 229544 & 2471122 & vibrobox                 & 12328  & 12328  & 342828  \\
\cline{5-8}
\cline{1-4}
\multicolumn{4}{c}{SAT14Literal}                                                                                  \\
\cline{1-4}
\cline{1-4}
6s133                           & 96430   & 140968 & 328924                                                         \\
6s153                           & 171292  & 245440 & 572692                                                         \\
aaai10-planning-ipc5            & 107838  & 308235 & 690466                                                         \\
atco\_enc2\_opt1\_05\_21        & 112732  & 526872 & 2097393                                                        \\
dated-10-11-u                   & 283720  & 629461 & 1429872                                                        \\
\bottomrule
\end{tabular}
        \end{table}
\vfill
\newpage

\begin{table}\caption{Basic properties of our large benchmark subset.}
\label{tbl:instanceslarge}

\centering
\begin{tabular}{@{\extracolsep{2pt}}lrrrlrrr@{}}
  \toprule
  Hypergraph                         & $n$                & $m$                & $p$                & Hypergraph               & $n$               & $m$                & $p$                 \\
  \cline{5-8}
\cline{1-4}
\multicolumn{4}{c}{DAC2012}      & \multicolumn{4}{c}{SAT14Primal}                                                                                                                        \\
\cline{1-4}
\cline{5-8}
superblue19                        & \numprint{522482}  & \numprint{511685}  & \numprint{1713796} & AProVE07-27              & \numprint{7729}   & \numprint{29194}   & \numprint{77124}    \\
superblue16                        & \numprint{698339}  & \numprint{697458}  & \numprint{2280417} & countbitssrl032          & \numprint{18607}  & \numprint{55724}   & \numprint{130020}   \\
superblue14                        & \numprint{630802}  & \numprint{619815}  & \numprint{2048903} & 6s184                    & \numprint{33365}  & \numprint{97516}   & \numprint{227536}   \\
superblue3                         & \numprint{917944}  & \numprint{898001}  & \numprint{3109446} & 6s9                      & \numprint{34317}  & \numprint{100384}  & \numprint{234228}   \\
\cline{1-4}
\multicolumn{4}{c}{ISPD98}       & 6s133              & \numprint{48215}   & \numprint{140968}  & \numprint{328924}                                                                       \\
\cline{1-4}
ibm09                              & \numprint{53395}   & \numprint{60902}   & \numprint{222088}  & 6s153                    & \numprint{85646}  & \numprint{245440}  & \numprint{572692}   \\
ibm11                              & \numprint{70558}   & \numprint{81454}   & \numprint{280786}  & atco\_enc1\_opt2\_10\_16 & \numprint{9643}   & \numprint{152744}  & \numprint{641139}   \\
ibm10                              & \numprint{69429}   & \numprint{75196}   & \numprint{297567}  & aaai10-planning-ipc5     & \numprint{53919}  & \numprint{308235}  & \numprint{690466}   \\
ibm12                              & \numprint{71076}   & \numprint{77240}   & \numprint{317760}  & hwmcc10-timeframe        & \numprint{163622} & \numprint{488120}  & \numprint{1138944}  \\
ibm13                              & \numprint{84199}   & \numprint{99666}   & \numprint{357075}  & itox\_vc1130             & \numprint{152256} & \numprint{441729}  & \numprint{1143974}  \\
ibm14                              & \numprint{147605}  & \numprint{152772}  & \numprint{546816}  & dated-10-11-u            & \numprint{141860} & \numprint{629461}  & \numprint{1429872}  \\
ibm15                              & \numprint{161570}  & \numprint{186608}  & \numprint{715823}  & atco\_enc1\_opt2\_05\_4  & \numprint{14636}  & \numprint{386163}  & \numprint{1652800}  \\
ibm16                              & \numprint{183484}  & \numprint{190048}  & \numprint{778823}  & manol-pipe-g10bid\_i     & \numprint{266405} & \numprint{792175}  & \numprint{1848407}  \\ 
ibm18                              & \numprint{210613}  & \numprint{201920}  & \numprint{819697}  & manol-pipe-c8nidw        & \numprint{269048} & \numprint{799867}  & \numprint{1866355}  \\
ibm17                              & \numprint{185495}  & \numprint{189581}  & \numprint{860036}  & atco\_enc2\_opt1\_05\_21 & \numprint{56533}  & \numprint{526872}  & \numprint{2097393}  \\
\cline{1-4}
\multicolumn{4}{c}{SAT14Dual}    & dated-10-17-u      & \numprint{229544}  & \numprint{1070757} & \numprint{2471122}                                                                      \\
\cline{1-4}
 AProVE07-27                       & \numprint{29194}   & \numprint{7729}    & \numprint{77124}   & ACG-20-5p0               & \numprint{324716} & \numprint{1390931} & \numprint{3269132}  \\
countbitssrl032                    & \numprint{55724}   & \numprint{18607}   & \numprint{130020}  & ACG-20-5p1               & \numprint{331196} & \numprint{1416850} & \numprint{3333531}  \\                       
\cline{5-8}
6s184                              & \numprint{97516}   & \numprint{33365}   & \numprint{227536}  & \multicolumn{4}{c}{SPM}                                                                 \\
\cline{5-8}
6s9                                & \numprint{100384}  & \numprint{34317}   & \numprint{234228}  & powersim                 & \numprint{15838}  & \numprint{15838}   & \numprint{67562}    \\
6s133                              & \numprint{140968}  & \numprint{48215}   & \numprint{328924}  & as-caida                 & \numprint{31379}  & \numprint{26475}   & \numprint{106762}   \\
6s153                              & \numprint{245440}  & \numprint{85646}   & \numprint{572692}  & hvdc1		     & \numprint{24842}  & \numprint{24842}   & \numprint{159981}   \\
atco\_enc1\_opt2\_10\_16           & \numprint{152744}  & \numprint{9643}    & \numprint{641139}  & Ill\_Stokes              & \numprint{20896}  & \numprint{20896}   & \numprint{191368}   \\
aaai10-planning-ipc5               & \numprint{308235}  & \numprint{53919}   & \numprint{690466}  & mult\_dcop\_01           & \numprint{25187}  & \numprint{25187}   & \numprint{193276}   \\
hwmcc10-timeframe                  & \numprint{488120}  & \numprint{163622}  & \numprint{1138944} & lp\_pds\_20              & \numprint{108175} & \numprint{33798}   & \numprint{232647}   \\
itox\_vc1130                       & \numprint{441729}  & \numprint{152256}  & \numprint{1143974} & lhr14		     & \numprint{14270}  & \numprint{14270}   & \numprint{307858}   \\
dated-10-11-u                      & \numprint{629461}  & \numprint{141860}  & \numprint{1429872} & c-61		     & \numprint{43618}  & \numprint{43618}   & \numprint{310016}   \\
atco\_enc1\_opt2\_05\_4            & \numprint{386163}  & \numprint{14636}   & \numprint{1652800} & ckt11752\_dc\_1          & \numprint{49702}  & \numprint{49702}   & \numprint{333029}   \\
manol-pipe-g10bid\_i               & \numprint{792175}  & \numprint{266405}  & \numprint{1848407} & RFdevice                 & \numprint{74104}  & \numprint{74104}   & \numprint{365580}   \\
manol-pipe-c8nidw                  & \numprint{799867}  & \numprint{269048}  & \numprint{1866355} & light\_in\_tissue        & \numprint{29282}  & \numprint{29282}   & \numprint{406084}   \\
atco\_enc2\_opt1\_05\_21           & \numprint{526872}  & \numprint{56533}   & \numprint{2097393} & Pres\_Poisson            & \numprint{14822}  & \numprint{14822}   &	\numprint{715804}   \\
dated-10-17-u			   & \numprint{1070757} & \numprint{229544}  & \numprint{2471122} & Andrews                  & \numprint{60000}  & \numprint{60000}   & \numprint{760154}   \\  
ACG-20-5p0			   & \numprint{1390931} & \numprint{324716}  & \numprint{3269132} & 2D\_54019\_highK         & \numprint{54019}  & \numprint{54019}   & \numprint{996414}   \\  
ACG-20-5p1			   & \numprint{1416850} & \numprint{331196}  & \numprint{3333531} & case39                   & \numprint{40216}  & \numprint{40216}   & \numprint{1042160}  \\  
\cline{1-4}
\multicolumn{4}{c}{SAT14Literal} & denormal           & \numprint{89400}   & \numprint{89400}   & \numprint{1156224}                                                                      \\ 
\cline{1-4}
AProVE07-27                        & \numprint{15458}   & \numprint{29194}   & \numprint{77124}   & 2cubes\_sphere           & \numprint{101492} & \numprint{101492}  & \numprint {1647264} \\        
countbitssrl032			   & \numprint{37213}	& \numprint{55724}   & \numprint{130020}  & av41092                  & \numprint{41092}  & \numprint{41092}   & \numprint{1683902}  \\  
6s184				   & \numprint{66730}	& \numprint{97516}   & \numprint{227536}  & Lin                      & \numprint{256000} & \numprint{256000}  & \numprint{1766400}  \\  
6s9                                & \numprint{68634}   & \numprint{100384}  & \numprint{234228}  & cfd1                     & \numprint{70656}  & \numprint{70656}   & \numprint{1828364}  \\   
6s133                              & \numprint{96430}   & \numprint{140968}  & \numprint{328924}  & opt1                     & \numprint{15449}  & \numprint{15449}   & \numprint{1930655}  \\ 
6s153                              & \numprint{171292}  & \numprint{245440}  & \numprint{572692}  & mixtank\_new             & \numprint{29957}  & \numprint{29957}   &	\numprint{1995041}  \\ 
atco\_enc1\_opt2\_10\_16           & \numprint{18930}   & \numprint{152744}  & \numprint{641139}  & sme3Db                   & \numprint{29067}  & \numprint{29067}   & \numprint{2081063}  \\ 
aaai10-planning-ipc5		   & \numprint{107838}	& \numprint{308235}  & \numprint{690466}  & mc2depi                  & \numprint{525825} & \numprint{525825}  & \numprint{2100225}  \\ 
hwmcc10-timeframe		   & \numprint{327243}	& \numprint{488120}  & \numprint{1138944} & poisson3Db               & \numprint{85623}  & \numprint{85623}   & \numprint{2374949}  \\ 
itox\_vc1130			   & \numprint{294326}	& \numprint{441729}  & \numprint{1143974} & rgg\_n\_2\_18\_s0        & \numprint{262144} & \numprint{262141}  & \numprint{3094566}  \\ 
dated-10-11-u			   & \numprint{283720}	& \numprint{629461}  & \numprint{1429872} & cnr-2000                 & \numprint{325557} & \numprint{247501}  & \numprint{3216152}  \\ 
atco\_enc1\_opt2\_05\_4		   & \numprint{28738}	& \numprint{386163}  & \numprint{1652800} & m14b                     & \numprint{214765} & \numprint{214765}  & \numprint{3358036}  \\ 
manol-pipe-g10bid\_i		   & \numprint{532810}	& \numprint{792175}  & \numprint{1848407} & laminar\_duct3D          & \numprint{67173}  & \numprint{67173}   & \numprint{3833077}  \\ 
manol-pipe-c8nidw		   & \numprint{538096}	& \numprint{799867}  & \numprint{1866355} & gearbox                  & \numprint{153746} & \numprint{153746}  &	\numprint{9080404}  \\ 
atco\_enc2\_opt1\_05\_21	   & \numprint{112732}	& \numprint{526872}  & \numprint{2097393} & BenElechi1               & \numprint{245874} & \numprint{245874}  & \numprint{13150496} \\ 
dated-10-17-u			   & \numprint{459088}	& \numprint{1070757} & \numprint{2471122} & af\_shell1               & \numprint{504855} & \numprint{504855}  & \numprint{17588875} \\ 
ACG-20-5p0			   & \numprint{649432}	& \numprint{1390931} & \numprint{3269132}                                                                                           \\
ACG-20-5p1			   & \numprint{662392}	& \numprint{1416850} & \numprint{3333531}                                                                                           \\
\bottomrule
\end{tabular}
\end{table}
\vfill
\newpage

\end{appendix}

\end{document}